\documentclass[aps,prl,twocolumn,10pt,amsmath,superscriptaddress,amssymb,longbibliography]{revtex4-1}
\usepackage{epsfig}
\usepackage{epstopdf}
\usepackage{dcolumn}
\usepackage{amsbsy}
\usepackage{bm}
\usepackage{color,CJK}
\usepackage{amsmath,amssymb,amsfonts}
\usepackage{appendix}
\usepackage{graphicx}
\usepackage{algorithm}
\usepackage{enumerate}
\usepackage{makeidx}
\usepackage{braket}
\usepackage[usenames,dvipsnames]{xcolor}
\usepackage[driverfallback=dvipdfmx,colorlinks,linkcolor=blue,citecolor=blue,urlcolor=blue]{hyperref}

\begin{document}

\title{Experimental realization of spin-tensor momentum coupling in ultracold Fermi gases}

\author{Donghao Li}
\affiliation{State Key Laboratory of Quantum Optics and Quantum
Optics Devices, \\ Institute of Opto-electronics, Shanxi University,
Taiyuan, Shanxi 030006, People's Republic of China }
\affiliation{Collaborative Innovation Center of Extreme Optics,
Shanxi University, Taiyuan, Shanxi 030006, People's Republic of
China }

\author{Lianghui Huang}
\email[Corresponding author email: ]{huanglh06@126.com; }
\affiliation{State Key Laboratory of Quantum Optics and Quantum
Optics Devices, \\ Institute of Opto-electronics, Shanxi University,
Taiyuan, Shanxi 030006, People's Republic of China }
\affiliation{Collaborative Innovation Center of Extreme Optics,
Shanxi University, Taiyuan, Shanxi 030006, People's Republic of
China }

\author{Peng Peng}
\affiliation{State Key Laboratory of Quantum Optics and Quantum
Optics Devices, \\ Institute of Opto-electronics, Shanxi University,
Taiyuan, Shanxi 030006, People's Republic of China }
\affiliation{Collaborative Innovation Center of Extreme Optics,
Shanxi University, Taiyuan, Shanxi 030006, People's Republic of
China }

\author{Guoqi Bian}
\affiliation{State Key Laboratory of Quantum Optics and Quantum
Optics Devices, \\ Institute of Opto-electronics, Shanxi University,
Taiyuan, Shanxi 030006, People's Republic of China }
\affiliation{Collaborative Innovation Center of Extreme Optics,
Shanxi University, Taiyuan, Shanxi 030006, People's Republic of
China }

\author{Pengjun Wang}
\affiliation{State Key Laboratory of Quantum Optics and Quantum
Optics Devices, \\  Institute of Opto-electronics, Shanxi
University, Taiyuan, Shanxi 030006, People's Republic of China }
\affiliation{Collaborative Innovation Center of Extreme Optics,
Shanxi University, Taiyuan, Shanxi 030006, People's Republic of
China }

\author{Zengming Meng}
\affiliation{State Key Laboratory of Quantum Optics and Quantum
Optics Devices, \\ Institute of Opto-electronics, Shanxi University,
Taiyuan, Shanxi 030006, People's Republic of China }
\affiliation{Collaborative Innovation Center of Extreme Optics,
Shanxi University, Taiyuan, Shanxi 030006, People's Republic of
China }

\author{Liangchao Chen}
\affiliation{State Key Laboratory of Quantum Optics and Quantum
Optics Devices, \\  Institute of Opto-electronics, Shanxi
University, Taiyuan, Shanxi 030006, People's Republic of China }
\affiliation{Collaborative Innovation Center of Extreme Optics,
Shanxi University, Taiyuan, Shanxi 030006, People's Republic of
China }

\author{Jing Zhang}
\email[Corresponding author email: ]{jzhang74@yahoo.com;
\\jzhang74@sxu.edu.cn}
\affiliation{State Key Laboratory of Quantum Optics and Quantum
Optics Devices, \\  Institute of Opto-electronics, Shanxi
University, Taiyuan, Shanxi 030006, People's Republic of China }

\date{\today }

\begin{abstract}
We experimentally realize the spin-tensor momentum coupling (STMC) using the three ground Zeeman states coupled by three Raman laser beams in ultracold atomic system of $^{40}$K Fermi atoms. This new type of STMC consists of two bright-state bands as a
regular spin-orbit coupled spin-1/2 system and one dark-state middle band. Using radio-frequency spin-injection spectroscopy, we investigate the energy band of STMC. It is demonstrated that the middle state is a dark state in the STMC system. The realized energy band of STMC may open the door for further exploring exotic quantum matters.
\end{abstract}

\maketitle
Ultracold atomic gases provide a versatile platform for exploring many interesting quantum phenomena~\cite{RMP1,RMP2,RMP3,RMP4}, which give insights
into systems that are difficult to realize in solid state systems~\cite{RMP6,RMP7,RMPQi}, and especially study quantum matter
in the presence of a variety of gauge fields~\cite{RMP8, RMP9, Goldman2014, Zhai2015,Zhang2014,Zhu2018}. A prominent example is the spin-orbit coupling (SOC), which is responsible for fascinating
phenomena such as topological insulators and superconductors~\cite{RMP7,RMPQi}, quantum spin Hall effect~\cite{RMP11}. The
synthetic one-dimensional (1D) SOC generated by a Raman transition has been implemented experimentally for bosonic~\cite{RMP13} and
fermionic~\cite{RMP14,RMP15} atoms. The 1D SOC has also been realized with lanthanide and alkali earth
atoms~\cite{RMP16,RMP18,RMP19}. Recently, the experimental realizations of two-dimensional SOC
have been respectively reported in ultracold Fermi gases of $^{40}$K~\cite{RMP20,RMP21} using a tripod scheme in a continuum space
and Bose-Einstein condensate (BEC) of $^{87}$Rb~\cite{RMP22} using a scheme called optical Raman lattice in two-dimensional Brillouin zone, where the Dirac point and nontrivial band topology are observed. All of these proposed and realized various types focus on spin-vector momentum coupling for both spin 1/2 and
1~\cite{RMP13,RMP14,RMP15,RMP20,RMP21,RMP22,Fu2011,Shuai2012,Qu2013,Olson2014}, while high-order spin-tensors naturally exist in a high-spin (larger or equal to 1) system.

A theoretical scheme for realizing spin-tensor momentum coupling (STMC) of spin-1 atoms has been proposed recently and some interesting phenomena were predicted~\cite{RMP26}. Here, STMC consists of two bright-state bands as a
regular spin-orbit coupled spin-1/2 system and one dark-state middle band. The middle-band minimum is close to that of two bright-states, so significantly modifies the state density of system ground state. This effect combining with interaction can offer a possible way to generate a new type of dynamical stripe states with high visibility and long tunable
periods~\cite{RMP26}, so can bring the advantage for the direct experimental observation. Furthermore, the more complex spin-tensor momentum coupling~\cite{RMP27}, such as, can induce different types of triply degenerate points connected by intriguing
Fermi arcs at surfaces. Therefore the STMC changes the band structure dramatically, and leads to many interesting many-body physics in the presence of interactions between atoms. In this paper, we experimentally realize this new type of STMC with two bright-state bands and one dark-state middle band in spin-1 ultracold Fermi gases based on the scheme in Ref.~\cite{RMP26}.

Dark states in quantum optics~\cite{Scully-book} and atom optics~\cite{Arimondo1996} are well studied and have led to electromagnetically induced transparency (EIT)~\cite{Boller1991,Fleischhauer2005}, stimulated Raman adiabatic passage (STIRAP)~\cite{Kuklinski1989} and subrecoil cooling scheme such as velocity selective coherent population trapping (VSCPT)~\cite{Aspect1988}. Dark states are superpositions of internal atomic ground states which are decoupled from coupling and have not energy shift induced by coupling. In contrast, bright-states have energy shift depending on coupling strength. For example, considering a lambda atomic systems (two ground states and one excited state) coupled with a pair of near-resonant fields, the excitation amplitudes of different ground states to the same excited state destructively interfere to generate a dark state. Thus when an atom is populated in such a dark state, it remains unexcited and cannot fluoresce. In this paper, we study STMC with the bright- and dark-states in the continuous momentum space.

\begin{figure}[!htb]
\includegraphics[width=3.2in]{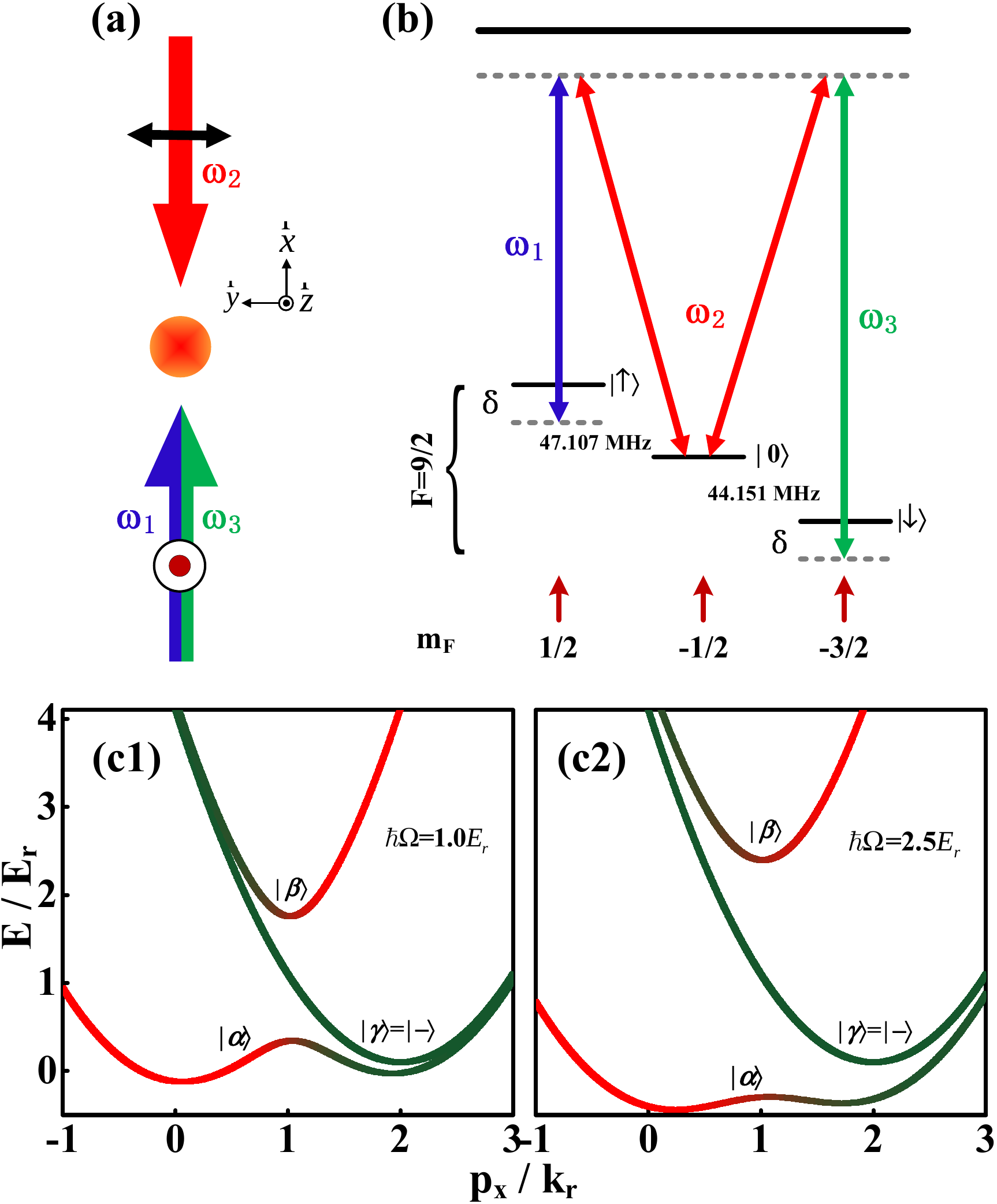}
\caption{(Color online) \textbf{Schematics of the Raman lasers configuration and atomic levels of generating STMC.}
(a)~Raman lasers configuration to generate STMC in ultracold Fermi gases.
(b)~Raman transitions between three hyperfine spin states with detuning $\delta$.
(c1) and (c2)~Theoretical single-particle band structure for Raman strength $\hbar \Omega_{R}$=$1.0E_{r}$ and $2.5E_{r}$ respectively. The detuning $\hbar \delta$ is set as $0.1E_{r}$. The lowest band indicate the eigenstate $\left\vert \alpha \right\rangle$, the hightest band indicate the eigenstate $\left\vert \beta \right\rangle$, and the middle one indicates the eigenstate $\left\vert \gamma \right\rangle$. }
\label{Fig1}
\end{figure}

The realization of STMC in ultracold Fermi gases of $^{40}$K atoms is illustrated in Fig.~\ref{Fig1}(a), which is similar with the scheme~\cite{RMP26}. We choose three magnetic sublevels $\left\vert \uparrow
\right\rangle$ = $\left\vert F=9/2,m_{F}=1/2 \right\rangle$ ($\left\vert 9/2,1/2 \right\rangle$), $\left\vert 0 \right\rangle$ = $\left\vert 9/2,-1/2 \right\rangle$, and  $\left\vert \downarrow \right\rangle$ = $\left\vert 9/2,-3/2 \right\rangle$ of the $F=9/2$ hyperfine level of $^{40}$K atomic electronic ground state as the three internal spin states, where $F$ denotes the total spin and $m_F$ is the magnetic quantum number. The three spin states are coupled by three Raman laser beams to generate STMC as shown in Fig. 1(a) and (b). Here, two of the laser beams 1, 3 and third laser 2 oppositely propagate along
$\hat{x}$ direction. Therefore, the three lasers induce two Raman transitions between hyperfine spin states $\left\vert 0 \right\rangle$ to $\left\vert \uparrow (\downarrow)\right\rangle$ state with coupling strength $\Omega_{ij}$, both of which have the same recoil momentum 2$\hbar k_{r}$ along the $\hat{x}$ direction. Two Raman couplings flip atoms from $\left\vert 0 \right\rangle$ to $\left\vert \uparrow (\downarrow)\right\rangle$ spin states and simultaneously impart momentum 2$\hbar k_{r}$ via the two-photon Raman process. However, the two spin states
$\left\vert \uparrow \right\rangle$ and $\left\vert \downarrow \right\rangle$ aren't coupled via Raman process due to $\Delta m_F >1$ as shown in Fig.~\ref{Fig1}(b). The single particle motion along $\hat{x}$ direction can be expressed as the STMC Hamiltonian,

\emph{\begin{equation}\label{1}
  H= \hbar\left(
      \begin{array}{ccc}
\frac{\hbar\mathbf{p}^2}{2m}+\delta & -\frac{\Omega_{12}}{2}e^{i2k_{r}x} & 0 \\
        -\frac{\Omega_{12}}{2}e^{-i2k_{r}x} &\frac{\hbar\mathbf{p}^2}{2m} & -\frac{\Omega_{23}}{2}e^{i2k_{r}x} \\
        0 & -\frac{\Omega_{23}}{2}e^{-i2k_{r}x} &\frac{\hbar\mathbf{p}^2}{2m}+\delta \\
      \end{array}
    \right).
\end{equation}}
Here, $\delta$ is the two-photon Raman detuning, $\hbar k_{r}$ is the single-photon
 recoil momentum of the Raman lasers, $\Omega_{ij}$ is the coupling strength between the state $\left\vert i \right\rangle$ and
 $\left\vert j \right\rangle$~\cite{RMP29}, and $\hbar$ is the Planck's constant. In order to eliminate the space dependence of the off-diagonal terms for Raman coupling in the original
Hamiltonian, one can apply an unitary transformation

\emph{\begin{equation}\label{2}
  U= \left(
      \begin{array}{ccc}
        e^{-i2k_{r}x} & 0 & 0 \\
        0 & 1 & 0 \\
        0 & 0 & e^{-i2k_{r}x} \\
      \end{array}
    \right),
\end{equation}}
to get the effective Hamiltonian

\emph{\begin{eqnarray}\label{eq:H3}
  H_{eff}=\hbar\left(
      \begin{array}{ccc}
        \frac{\hbar(p_{x}-2k_{r})^2}{2m}+\delta & -\frac{\Omega}{2} & 0 \\
        -\frac{\Omega}{2} & \frac{\hbar{p}^{2}_{x}}{2m} & -\frac{\Omega}{2} \\
        0 & -\frac{\Omega}{2} & \frac{\hbar(p_{x}-2k_{r})^2}{2m}+\delta \\
      \end{array}
    \right) \nonumber\\
    =\frac{\hbar^2 p_{x}^2}{2m}+(\delta+\frac{2\hbar^2 k_{r}^2}{m}-\frac{2\hbar^2 k_{r}p_{x}}{m})F_{z}^{2}-\frac{\Omega}{2}F_{x}.
\end{eqnarray}}
Here we set $\Omega_{12}=\Omega_{23}=\Omega$, $p_{x}$ indicates the quasimomentum along $x$ direction. Here, a spin-1 system is spanned by nine basis operators, which include the identity operator (I), the three vector spin operators ($F_{x}$, $F_{y}$, and $F_{z}$) and the five spin quadrupole operators~\cite{Marti2016}. The operators $F_{x}$ and $F_{z}$ can be written in the matrix form
\emph{\begin{eqnarray}\label{eq:F}
  F_{x}=\left(
      \begin{array}{ccc}
        0& 1 & 0 \\
      1& 0 & 1 \\
        0 & 1 & 0 \\
      \end{array}
    \right), \nonumber
          F_{z}=\left(
      \begin{array}{ccc}
        1& 0 & 0 \\
      0& 0 & 0 \\
        0 & 0 & -1 \\
      \end{array}
    \right). \nonumber \\
\end{eqnarray}}
The term $p_{x}F_{z}^{2}$ describes
the one-dimensional coupling between a spin tensor and the linear-momentum (i.e., the spin-tensor momentum
coupling). We define the recoil momentum $\hbar k_{r}=2\pi
 \hbar /\lambda$ and recoil energy $E_{r}=(\hbar k_{r})^{2}/2m = \hbar \Omega_{0} = h\times8.45$ kHz as the natural momentum and energy
 units, where $m$ is the atomic mass of $^{40}$K, and $\lambda=768.85$ nm is the wavelength of the Raman laser.

The three dressed eigenstates of Eq. \ref{eq:H3} are expressed by the spin-1 basis $(\left\vert \uparrow \right\rangle ,|0\rangle ,\left\vert \downarrow \right\rangle )$

\begin{align}
\label{eq:H4}
\left\vert \alpha \right\rangle = & a_{1}\left\vert \uparrow \right\rangle + b_{1}\left\vert 0 \right\rangle + c_{1}\left\vert \downarrow \right\rangle, \\
\label{eq:H5}
\left\vert \beta \right\rangle = & a_{2}\left\vert \uparrow \right\rangle + b_{2}\left\vert 0 \right\rangle + c_{2}\left\vert \downarrow \right\rangle, \\
\label{eq:H6}
\left\vert \gamma \right\rangle = & a_{3}\left\vert \uparrow \right\rangle + b_{3}\left\vert 0 \right\rangle + c_{3}\left\vert \downarrow \right\rangle. \\ \nonumber
\end{align}%
where $a_{1}=c_{1}=1/\sqrt{u^{2}+2}$, $b_{1}=-u/\sqrt{u^{2}+2}$, and $u=((4p_{x}-\delta-4)-\sqrt{(4p_{x}-\delta-4)^{2}+2\Omega^{2}})/\Omega$. $a_{2}=c_{2}=1/\sqrt{v^{2}+2}$, $b_{2}=-v/\sqrt{v^{2}+2}$, and $v=((4p_{x}-\delta-4)+\sqrt{(4p_{x}-\delta-4)^{2}+2\Omega^{2}})/\Omega$. $a_{3}=-c_{3}=1/\sqrt{2}$ and $b_{3}=0$. The $\left\vert \alpha \right\rangle$ and $\left\vert \beta \right\rangle$ are the lowest and highest energy dressed states respectively. The $\left\vert \gamma \right\rangle$ is middle energy dressed state.

We define the spin components $\left\vert 0 \right\rangle$ and $\left\vert \pm \right\rangle$ = $(1/\sqrt{2})(\left\vert \uparrow \right\rangle \pm \left\vert \downarrow \right\rangle$). The middle state $\left\vert
  \gamma \right\rangle$ corresponds to the spin dressed component $\left\vert - \right\rangle$. For a single-particle energy band structure, the lowest and highest bands of STMC are the bright dressed states, which are composed of three spin components $\left\vert 0 \right\rangle$, $\left\vert \uparrow\right\rangle$ and $\left\vert \downarrow\right\rangle$ and the amplitude of three spin components depend on the $\Omega$ and $\delta$. The energy shift of the lowest and highest bands of STMC depends on the coupling strength as shown in Fig. 1(c1) and (c2). The highest band of STMC moves to higher energy and the lowest band to lower energy as the coupling strength increases and the detuning $\delta$ is fixed. The lowest and highest bands behave as a regular spin-orbit coupled spin-1/2 system. However, the middle state
 ($\left\vert \gamma \right\rangle$) is independent on the $\Omega$ and $\delta$ from Eq.~(\ref{eq:H6}). So the middle state
 $\left\vert \gamma \right\rangle$ has not energy shift by Raman coupling, therefore called dark state. The dark-state band plays an important role on both ground-state and dynamical properties of the interacting BECs with SOC as described in Ref~\cite{RMP26}.

We start quantum degenerate gases of $^{40}$K atoms at the spin state $|9/2,9/2\rangle$ by sympathetic evaporative cooling to 1.5 $\mu$K with $^{87}$Rb atoms at the spin state $|2,2\rangle$ in the
quadrupole-Ioffe configuration (QUIC) trap, and then transport them into the center of the glass cell in favor of optical access, which is
used in previous experiments~\cite{RMP29,RMP30}. Subsequently, we typically get the degenerate Fermi gas of $(\sim 4\times 10^6)$ $^{40}$K
atoms in the lowest hyperfine Zeeman state $|9/2,9/2\rangle$ by gradually decreasing the depth of the optical trap. Finally, we obtain
ultracold Fermi gases with the temperature around $0.3$ $T_F$, where the Fermi temperature is defined by $T_F=\hbar \bar{\omega} (6N)^{1/3}/k_B$. Here $\bar{\omega} =(\omega_x\omega_y\omega_z)^{1/3}$ $\simeq2$$\pi\times80$ Hz is the geometric mean of the optical trap
frequency for $^{40}$K degenerate Fermi gas in our experiment, $N$ is the particle number of $^{40}$K atoms, and $k_B$ is the Boltzmann's
constant. After the evaporation, the remaining $^{87}$Rb atoms are removed by shining a resonant laser beam pulse (780 nm) for 0.03 $ms$
without heating and losing $^{40}$K atoms. Afterwards, the ultracold Fermi gases of $|9/2,9/2\rangle$ state in optical dipole trap are transferred into the spin state $|9/2,3/2\rangle$ using a rapid adiabatic passage induced by a rf field with duration of 80 ms at $B\simeq19.6$ G to avoid the atomic
loss because of other Feshbach resonance~\cite{RMP31,RMP32}, where the center frequency of RF field is 6.17 MHz and the
scanning width is 0.4 MHz.

Three laser beams of 768.85 nm are used as the Raman lasers to generate the STMC along $\hat{x}$, which are
extracted from a continuous-wave Ti:sapphire single frequency laser. The Raman beams 1 and 2 are frequency-shifted around 74.896 MHz and 122 MHz by
two single pass acousto-optic modulators (AOM), respectively. The Raman beams 3 is double pass frequency-shifted around $166.15$
MHz by AOM. Afterwards, the Raman beams 1 and 3 are coupled with the same polarization into one polarization maintaining single-mode fibers, and Raman beam 2 is sent to the second single-mode fiber to increase the stability of the beam pointing and the quality of the beam profile. The two Raman lasers
1 and 3 from the first fiber and Raman laser 2 from second fiber counter-propagate along the x axis, and are focused at the position of the atomic cloud with $1/e^{2}$ radii of 200 $\mu$m, larger than the Fermi radius 43 $\mu$m of the degenerate Fermi gas~\cite{Giorgini2008}, as shown in Fig.~\ref{Fig1}(a). The two Raman lasers 1, 3 and Raman laser 2 are linearly polarized along the z and y directions, respectively, corresponding to drive $\pi$ and $\sigma$ transitions respectively relative to the quantization axis z shown in Fig.~\ref{Fig1}(a).

A homogeneous magnetic bias field $B_{exp}$ is applied in the $z$ axis (gravity direction) by a
pair of quadrupole coils described in Ref.~\cite{RMP20}, which generates Zeeman splitting on the ground hyperfine state. We ramp the magnetic-field to an expected field $B_{exp}=160$ G during 30 ms, and increase the intensity of
the three Raman laser beams to desired value in 20 ms to generate the STMC in three sublevels $\left\vert 9/2,1/2 \right\rangle$, $\left\vert 9/2,-1/2 \right\rangle$, and $\left\vert 9/2,-3/2 \right\rangle$ of ultracold Fermi gases. Here, we employ the spin injection spectrum to measure the energy band structure. So we prepare STMC as the final empty state and the other state $|9/2,3/2\rangle$ as the initial state. A
Gaussian shape pulse of the rf field is applied for 450 $\mu$s to drive atoms from $|9/2,3/2\rangle$ to the final empty state with STMC~\cite{RMP14,RMP15,RMP20}. Following the spin injection process, the Raman lasers, the optical trap and the magnetic field are switched off abruptly, and a magnetic field gradient is applied in the first 10 ms during the first free expansion, which creates a spatial separation of different Zeeman states due to Stern-Gerlach
effect. At last, the atoms are imaged along the $\textit{z}$ direction after total 12 $ms$ free expansion, which gives the momentum distribution for each spin component. By counting the number of atoms in the expected state as a function of the momentum and rf frequency from the
absorption image, the energy-band structure can be obtained.

\begin{figure}[!htb]
\centerline{\includegraphics[width=2.0in]{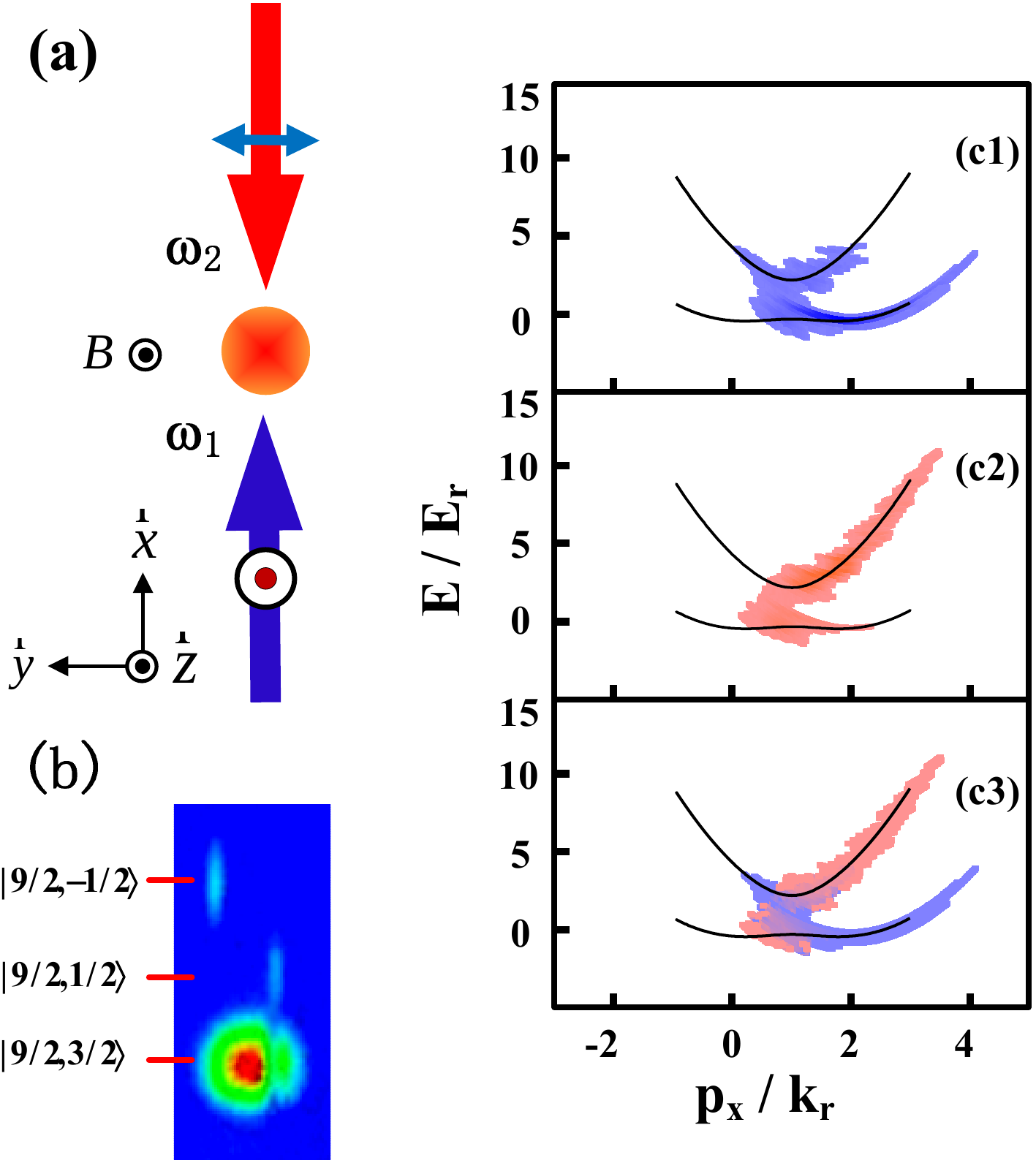}}
\caption{(Color online). \textbf{Energy band structure of 1D SOC ultracold Fermi gases.} (a) A pair of Raman beams couple two spin states $|9/2,1/2\rangle$ ($\left\vert \uparrow \right\rangle$) and $|9/2,-1/2\rangle$ ($\left\vert 0 \right\rangle$) to generate 1D SOC system with the Raman coupling strength $\hbar \Omega_{R}$=2.5$E_{r}$
and the detuning $\hbar \delta$=0$E_{r}$. (b) Time-of-flight absorption image of spin injection spectroscopy at a given frequency of rf field. (c1) and (c2) Reconstructed momentum- and spin-resolved $\left\vert \uparrow \right\rangle$ (blue) and $\left\vert 0 \right\rangle$ (red) spectra, respectively, when driving atoms from the free spin state $|9/2,3/2\rangle$. (c3) Displaying two graphs (c1) and (c2) simultaneously.
 \label{Fig2} }
\end{figure}

 First, we measure energy-band structure of standard 1D SOC as shown in Fig.~\ref{Fig2} which is similar as that reported in Ref~\cite{RMP15}. We prepare the atoms in the free spin state $|3/2,9/2\rangle$, then switch on two raman lasers to generate the 1D SOC system with two spin states $\left\vert \uparrow \right\rangle$ and $\left\vert 0 \right\rangle$. Using rf spin injection, we get the energy-band structure of 1D SOC, which is agree with the theoretical calculation well as shown in Fig.~\ref{Fig2}(c).

\begin{figure*}[!htb]
\centerline{\includegraphics[width=6.0in]{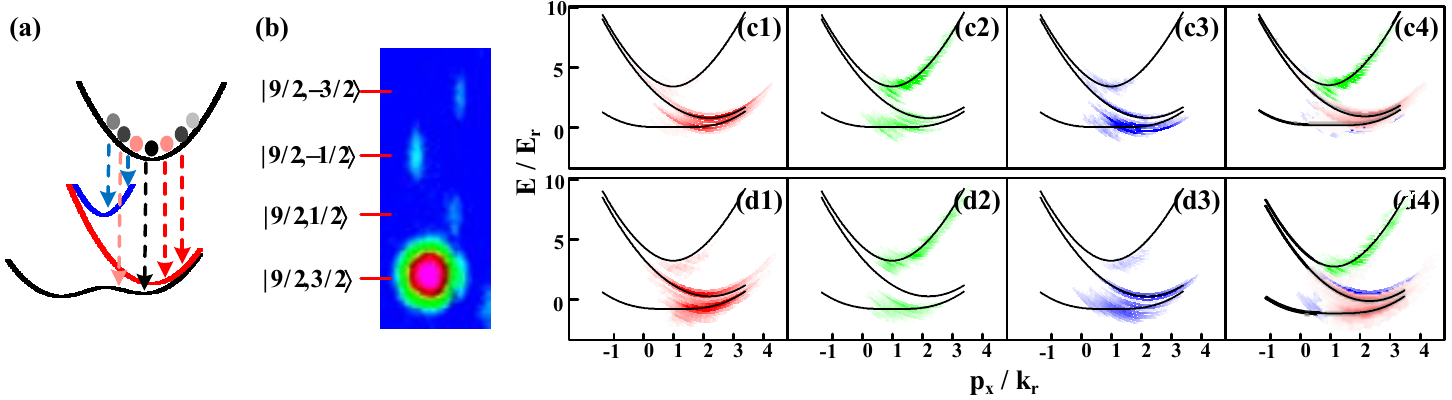}}
\caption{(Color online). \textbf{Energy band structure of STMC ultracold Fermi gases.} (a)~Schematic diagram of the process of spin injection spectroscopy with preparing in the free spin state $|9/2,3/2\rangle$. (b)~TOF image of spin injection spectroscopy at a given frequency of rf field. (c1)-(c3) Reconstructed spin-resolved $\left\vert \uparrow \right\rangle$ (red), $\left\vert 0 \right\rangle$ (green), and $\left\vert \downarrow \right\rangle$ (blue) spectra, respectively, when driving atoms from the free spin state $|9/2,3/2\rangle$ to the STMC system with the Raman coupling strength $\hbar \Omega$=2.5$E_{r}$ and the detuning $\hbar \delta$=0$E_{r}$. (c4) Displaying the three graphs (c1)-c3) simultaneously. (d1)-(d4) spin injection spectroscopy with the value of $\hbar \Omega$=3$E_{r}$ and $\hbar \delta$=0$E_{r}$. \label{Fig3} }
\end{figure*}

Now we study STMC and illustrate the middle state $\left\vert \gamma \right\rangle$ as a dark state in the STMC system. We prepare the ultracold atomic sample in the free state $|9/2,3/2\rangle$ with fixed magnetic field, then switch on three Raman laser to generate STMC system. Afterwards, we use rf
 spin injection from free state to empty STMC system as shown in Fig.~\ref{Fig3}(a). We obtain the
 energy spectrum of STMC system as shown in Fig.~\ref{Fig3}. Here, the detuning $\delta$=0, and the Raman coupling strength
 $\Omega$=2.5$\Omega_{0}$ (3$\Omega_{0}$) shown in Fig.~\ref{Fig3}(c) ((d)).

The three spin components appear in TOF images (For example in Fig.~\ref{Fig3}(b)) simultaneously when RF field drive atoms into the lowest and highest bands. The color depth contains the amplitude information of three spin components for the lowest and highest bands in Fig.~\ref{Fig3}(c) and (d). The highest band of STMC moving to higher energy and the lowest band to lower energy as the coupling strength increases are shown in Fig.~\ref{Fig3}(c) and (d). It illustrates that the lowest and highest bands are the bright dressed states and behave as a regular spin-orbit coupled spin-1/2 system. However, the middle dressed state $\left\vert \gamma \right\rangle$ only includes two spin components $\left\vert \uparrow\right\rangle$ and $\left\vert \downarrow\right\rangle$ (the spin dressed state $\left\vert - \right\rangle$). Therefore we only observe the two spin components $\left\vert \uparrow\right\rangle$ and $\left\vert \downarrow\right\rangle$ in the middle band from RF spectrum. Especially, almost no atoms in $\left\vert 0\right\rangle$ spin component are populated in the middle band as shown in Fig. 3(c2) and (d2). Moreover, Fig.~\ref{Fig3}(c) and (d) shows that the middle state $\left\vert \gamma \right\rangle$ is always a dark state without energy shift and decouples from the Raman strength $\Omega$.

\begin{figure}[!htb]
\centerline{\includegraphics[width=3.2in]{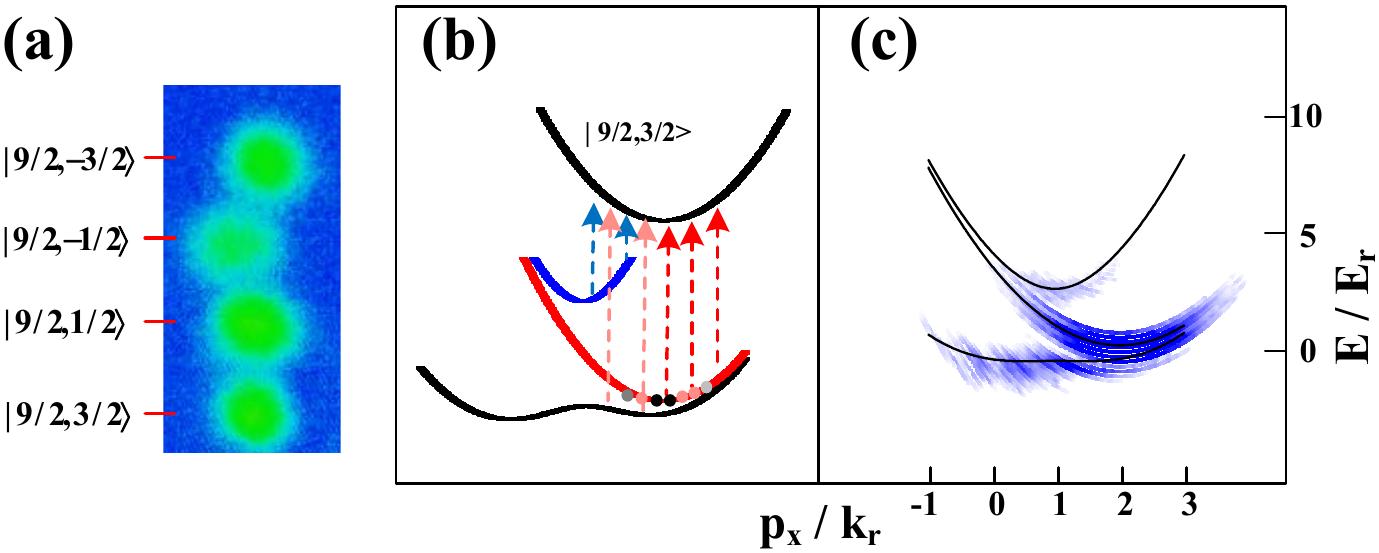}}
\caption{(Color online) \textbf{Another method to measuring Momentum-resolved rf spectroscopy of STMC ultracold Fermi gases.} (a)~Time-of-flight absorption image of rf spectroscopy at a given frequency of rf field. (b)~Schematic diagram of the process of rf spectroscopy with initially preparing atoms in STMC. (c) ~Reconstructed single particle dispersion and atom population when transferring atoms from the STMC ultracold Fermi gases system to free spin state $|9/2,3/2\rangle$ for $\hbar \Omega$=$2.5E_{R}$ and the detuning $\hbar \delta$=$0E_{R}$. \label{Fig4} }
\end{figure}

We also employ another RF spectrum method to measure energy-band structure of STMC state~\cite{RMP14}. Here, atoms are prepared in STMC as the initial state and the state $|9/2,3/2\rangle$ as the final empty state. We first prepare the ultracold atomic sample in the state $|9/2,1/2\rangle$ at first, then ramp on three Raman lasers with 5 ms to prepare Fermi atoms into STMC state in equilibrium. Then we apply rf pulse to drive the atoms from STMC state into free state $|9/2,3/2\rangle$ as shown in Fig.~\ref{Fig4}(a). We also get the energy spectrum of STMC system as shown in Fig.~\ref{Fig4}(b) and (c). Here, the Raman coupling strength $\Omega$=2.5$\Omega_{0}$, and the detuning $\delta$=0. For rf spectrum of STMC, the populated range into three bands of STMC by ramping on three Raman lasers is determined by the temperature of Fermi gases. The higher temperature of Fermi gases will make the momentum distribution broader, which will enlarge the measure range of energy band with compromising the signal-noise ratio of RF spectroscopy.

In conclusion, we have realized a scheme for generating STMC system in ultracold Fermi gases, and demonstrate the coupling between the spin tensors of atoms and their linear momenta. We measure and get the energy band structure of STMC via rf spin-injection spectrum. From rf spin-injection spectrum, we demonstrated that the middle state $\left\vert \gamma \right\rangle$ in the STMC system is a dark state. In this work, the dark-state band is not coupled with two bright-state bands through Raman coupling only for the single-particle picture. The dark-state band will couple with two bright-state bands in the presence of interactions between atoms. Thus, if we prepare atoms initially in the dark-state band, atoms will decay into the bright band due to interaction. Moreover, forming the dark-state band requires that the Raman detuning $\delta$ for $\left\vert \uparrow\right\rangle$ and $\left\vert \downarrow\right\rangle$ are exactly same. Otherwise, the dark-state band will change into the bright band. The experimental results may motivate more theoretical and experimental research of other novel topological matter, such as study super-solid like stripe orders due to the existence of the dark middle band, and may give rise to nontrivial topological matter with unprecedented properties in future.

\begin{acknowledgments}
We would like to thank Chuanwei Zhang for helpful discussions. This research was supported by the MOST (Grant No. 2016YFA0301602, 2018YFA0307601), NSFC (Grant No. 11474188, 11704234, 11904217), the Fund for Shanxi $\text{"1331 Project"}$ Key Subjects Construction, and the Program of Youth Sanjin Scholar.
\end{acknowledgments}

\end{document}